\documentclass[10pt,a4paper,english]{achemso}
\usepackage[T1]{fontenc}
\usepackage[latin9]{inputenc}
\usepackage{calc}
\usepackage{amsmath}
\usepackage{amssymb}
\usepackage{graphicx}
\usepackage{esint}

\makeatletter

\title{Direct signature of light-induced conical intersections in diatomics}
\author{G. J. Halász}
\affiliation{Department of Information Technology, University of Debrecen, H-4010
Debrecen, PO Box 12, Hungary}
\author{Á. Vibók}
\affiliation{Department of Theoretical Physics, University of Debrecen, H-4010
Debrecen, PO Box 5, Hungary}
\email{vibok@phys.unideb.hu}
\author{L. S. Cederbaum}
\affiliation{Theoretische Chemie, Physikalish-Chemisches Institut, Universität
Heidelberg, H-69120, Germany}
\pdfpageheight\paperheight
\pdfpagewidth\paperwidth

\newcommand{\noun}[1]{\textsc{#1}}

\usepackage{babel}

\usepackage{babel}

\makeatother

\usepackage{babel}
\begin{document}
\begin{abstract}
Nonadiabatic effects are ubiquitous in physics, chemistry and biology.
They are strongly amplified by conical intersections (CIs) which are
degeneracies between electronic states of triatomic or larger molecules.
A few years ago it has been revealed that CIs in molecular systems
can be formed by laser light even in diatomics. Due to the prevailing
strong nonadiabatic couplings, the existence of such laser-induced
conical intersections (LICIs) may considerably change the dynamical
behavior of molecular systems. By analyzing the photodissociation
process of the $\mathrm{D}_{2}^{+}$ molecule carefully, we found
a robust effect in the angular distribution of the photofragments
which serves as a direct signature of the LICI providing undoubted
evidence for its existence. 
\end{abstract}
\maketitle
\begin{center}
\framebox{\begin{minipage}[t]{1\columnwidth}%
\begin{center}
{}\includegraphics[width=0.3\paperwidth]{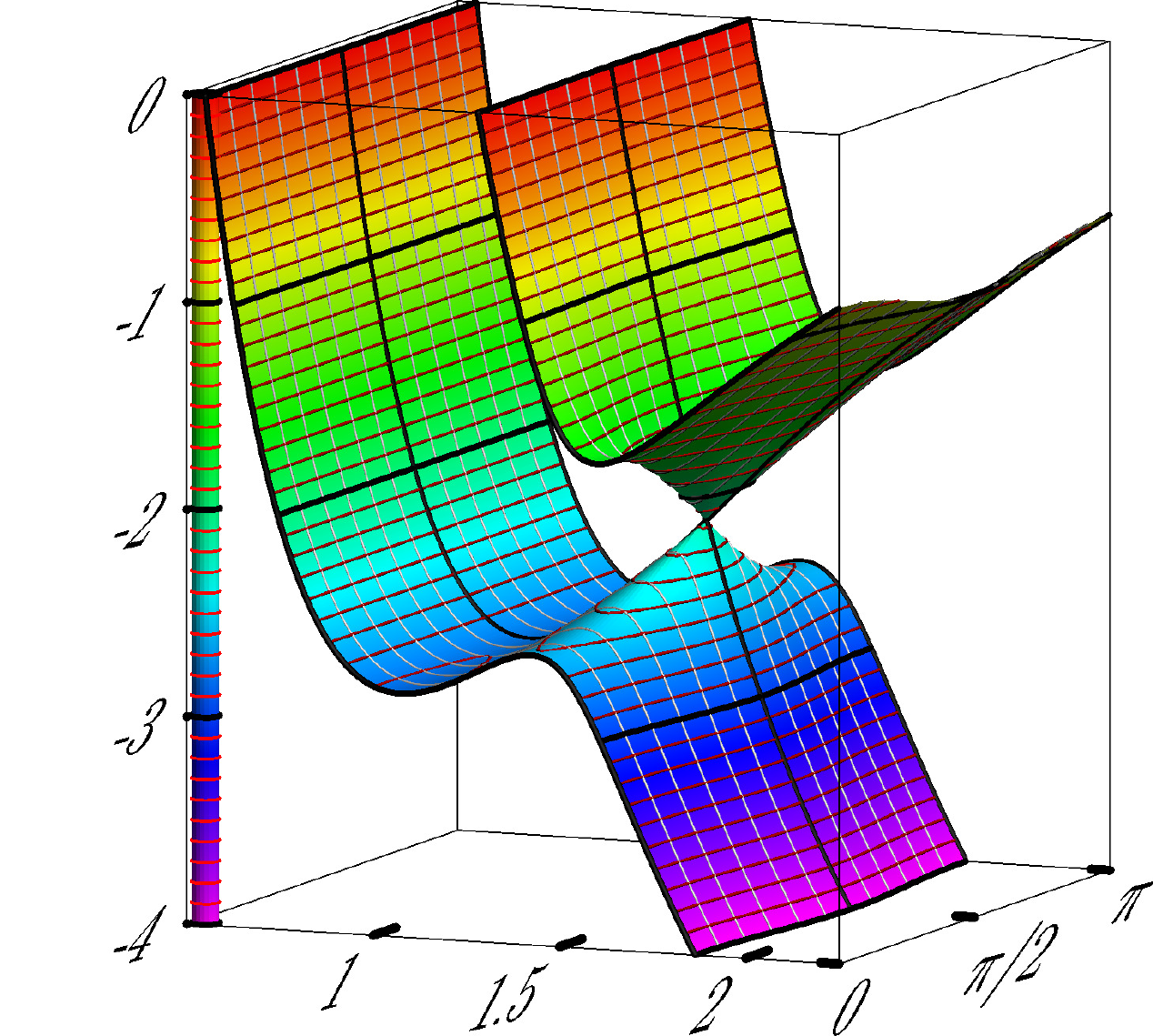} 
\par\end{center}

\begin{center}
{For Table of Contents Only.} 
\par\end{center}%
\end{minipage}}
\par\end{center}

It is well known that conical intersections (CIs) are widely recognized
to be ubiquitous in polyatomics, and to play an important role in
several different fields like spectroscopy, chemical reaction dynamics,
photophysics, photochemistry \cite{Zener,Teller1,Horst1,Truhlara,Baer1,Graham1,Wolfgang1,Baer2,Matsika1,Ben-Nur1,Andrzej1,Andrzej2,Ashfold1,Kim,Polli1,Truhlar1,Todd2,corkum,margaret1,margaret2}.
At CIs the nuclear and electronic motions couple strongly, giving
rise to often unexpected \textemdash{} so-called nonadiabatic \textemdash{}
phenomena. The nonadiabatic coupling matrix elements between electronic
states are the largest possible at CIs where they become singular
\cite{Horst1}. Therefore, one expects that such intersections give
rise to the strongest possible nonadiabatic phenomena. It was demonstrated
for a variety of examples that conical intersections provide the mechanism
for ultrafast chemical processes as, e.g. photodissociation, photoisomerization
and internal conversion to the electronic ground state \cite{Matsika1,Ben-Nur1,Andrzej1,Andrzej2,Ashfold1,Kim,Polli1,Truhlar1,Todd2,corkum,margaret1,margaret2}.
Clearly, the CIs must be considered as photochemically relevant decay
channels. The presence of CIs in biomolecules, like the building blocks
of DNA and proteins contribute to the photostability of these important
molecules. CIs can be formed between different electronic states starting
from triatomic systems to truly large polyatomic molecules. Several
important books, review articles and publications have documented
the existence and relevance of such intersections \cite{Zener,Teller1,Horst1,Truhlara,Baer1,Graham1,Wolfgang1,Baer2,Matsika1,Ben-Nur1,Andrzej1,Andrzej2,Ashfold1,Kim,Polli1,Truhlar1,Todd2,corkum,margaret1,margaret2}. 

CIs can exist only if the molecular system possesses at least two
independent nuclear degrees of freedom. This is why diatomics which
have only one nuclear vibrational coordinate cannot exhibit a CI.
Importantly, this statement only holds in field free space, a fact
which has been generally overlooked. When a laser field is present,
then due to the interaction of the diatomic with this field, the rotational
degree of freedom comes into play and serves as an additional degree
of freedom. It was revealed in earlier studies \cite{Nimrod1,Milan1}
that CIs can be induced both by running or standing laser waves even
in diatomics. Interestingly, in the standing wave case also the position
of the center of mass becomes a new degree of freedom in addition
to the nuclear vibrational and rotational motions. In order to demonstrate
how the rotation supplies the necessary degree of freedom to facilitate
the shaping of a CI in the general case of propagating laser waves,
one may adopt the dressed state representation \cite{floquet-1,floquet-2}.
In this picture the molecule-light interaction is explicitly included
into the Hamiltonian, and the changes of nuclear dynamics due to the
light field can be considered as originating from the appearance of
a \textquotedblleft{}light-induced conical intersection\textquotedblright{}
(LICI) \cite{Milan1}. The laser frequency determines the position
of the LICI, while the laser intensity determines the strength of
the nonadiabatic couplings. Our results in the last few years undoubtedly
demonstrated that LICIs exert strong effects on the quantum dynamics
even for weak laser fields \cite{Milan1,Gabor1,Gabor2,Gabor3,Gabor 4,Gabor 5,Gabor6,Gabor7,Lenz1}. 

The present study goes beyond previous investigations and makes an
attempt to provide and analyze a physical event which may serve as
an undoubted evidence of the laser- or light-induced conical intersection
(LICI), giving a \textquotedblleft{}direct signature\textquotedblright{}
of the presence of this intersection. It is known in the field of
nonadiabatic molecular dynamics that due to the extreme breakdown
of the Born-Oppenheimer approximation, conical intersections are responsible
for ultrafast radiationless processes, typically on the femtosecond
time scale. They provide pathways for extremely fast population transfer
between electronic states. This latter effect is probably the most
important inherent feature of the CIs. Nevertheless, until now one
could not find an unambiguous experimentally measurably quantity which
reflects directly this population transfer between electronic states
for a LICI. The present work discusses a physical process, where an
ultrafast population transfer takes place between the electronic states
of a diatomic molecule, providing direct evidence for the existence
of the LICI. The photodissociation process of the D$_{2}^{+}$ molecule
serves as a show case physical example. This molecule has already
been studied in vast amount of works \cite{Schumacher1,Bandrauk1,Bandrauk2,Bandrauk3,Charron1,Bandrauk4,Charron2,Atabek1,Sanding,Atabek2,Posthumus,Atabek3,Uhlmann1,Wang1,Anis1,Anis2,Esry1,Agnes1,Paul1,Thum1,Fischer1,Fischer2,Anis3,He1,Furakawa,Hendt1},
mainly because of its simplicity. It has the advantage that we can
compute it accurately and easily study the light-induced nonadiabatic
phenomena separately from other phenomena. 

It is important to note that recently, motivated by theoretical predictions
on the LICIs in diatomics, experiments on the laser-induced isomerization
and photodissociation of polyatomic molecules were qualitatively interpreted
using the concept of LICIs \cite{Buksbaum1,Banares1}. Furthermore,
theory devoted to generalize the LICI phenomenon for polyatomics has
been derived showing the large potential of LICIs in controlling reactions
\cite{Philip1}. 

\textbf{Description of the system.} The two relevant electronic states
of the D$_{2}^{+}$ ion (see Fig. 1), which will be considered in
the calculations are the ground (V$_{1}=1s\sigma_{g}$) and the first
excited (V$_{2}=2p\sigma_{u}$) eigenstates of the field-free Hamiltonian.
For describing the dissociation mechanism we assume that initially
the D$_{2}^{+}$ ion is in its ground electronic ($1s\sigma_{g}$)
as well as in its ground rotational state and in one of its vibrational
eigenstates (see Fig. 1). Exciting the electronic ground state by
a resonant laser pulse to the repulsive $2p\sigma_{u}$ state, the
two electronic states are resonantly coupled. The nonvanishing dipole
matrix elements are responsible for the light-induced electronic transitions.
Within these two electronic states representation the total time-dependent
Hamiltonian for the rovibronic nuclear motion reads

\begin{align}
H & =\left(\begin{array}{cc}
-\frac{1}{2\mu}\frac{\partial^{2}}{\partial R^{2}}+\frac{L_{\theta\varphi}^{2}}{2\mu R^{2}} & \;0\\
0 & \;-\frac{1}{2\mu}\frac{\partial^{2}}{\partial R^{2}}+\frac{L_{\theta\varphi}^{2}}{2\mu R^{2}}
\end{array}\right)+\label{eq:Hamilton}\\
 & \left(\begin{array}{cc}
V_{1}(R) & -\epsilon_{0}f(t)d(R)\cos\theta\cos\omega_{L}t\\
-\epsilon_{0}f(t)d(R)\cos\theta\cos\omega_{L}t & V_{2}(R)
\end{array}\right).\nonumber 
\end{align}
Here, R and ($\theta,\varphi$) are the molecular vibrational and
rotational coordinates, respectively, $\mu$ is the reduced mass,
and $L_{\theta\varphi}$ denotes the angular momentum operator of
the nuclei. Here $\theta$ is the angle between the polarization direction
and the direction of the transition dipole and thus one of the angles
of rotation of the molecule. $V_{1}(R)$ ($1s\sigma_{g}$) and $V_{2}(R)$
($2p\sigma_{u}$) are the potential energies of the two electronic
states coupled by the laser (whose frequency is $\omega_{L}$ and
amplitude is $\epsilon_{0}$), $f(t)$ is the envelop function and
$d(R)$$\left(=-\left\langle \psi_{1}^{e}\left|\sum_{j}r_{j}\right|\psi_{2}^{e}\right\rangle \right)$
is the transition dipole matrix element ($e=m_{e}=\hbar=1;$ atomic
units are used throughout the article). We used the quantities $V_{1}(R)$
and $V_{2}(R)$ and $d(R)$ published in \cite{dipol,pot}. 

\begin{figure*}[p]
\includegraphics[bb=35bp 0bp 365bp 328bp,clip,width=0.43\textwidth]{pot-3d-a}\hspace*{\fill}\includegraphics[width=0.54\textwidth]{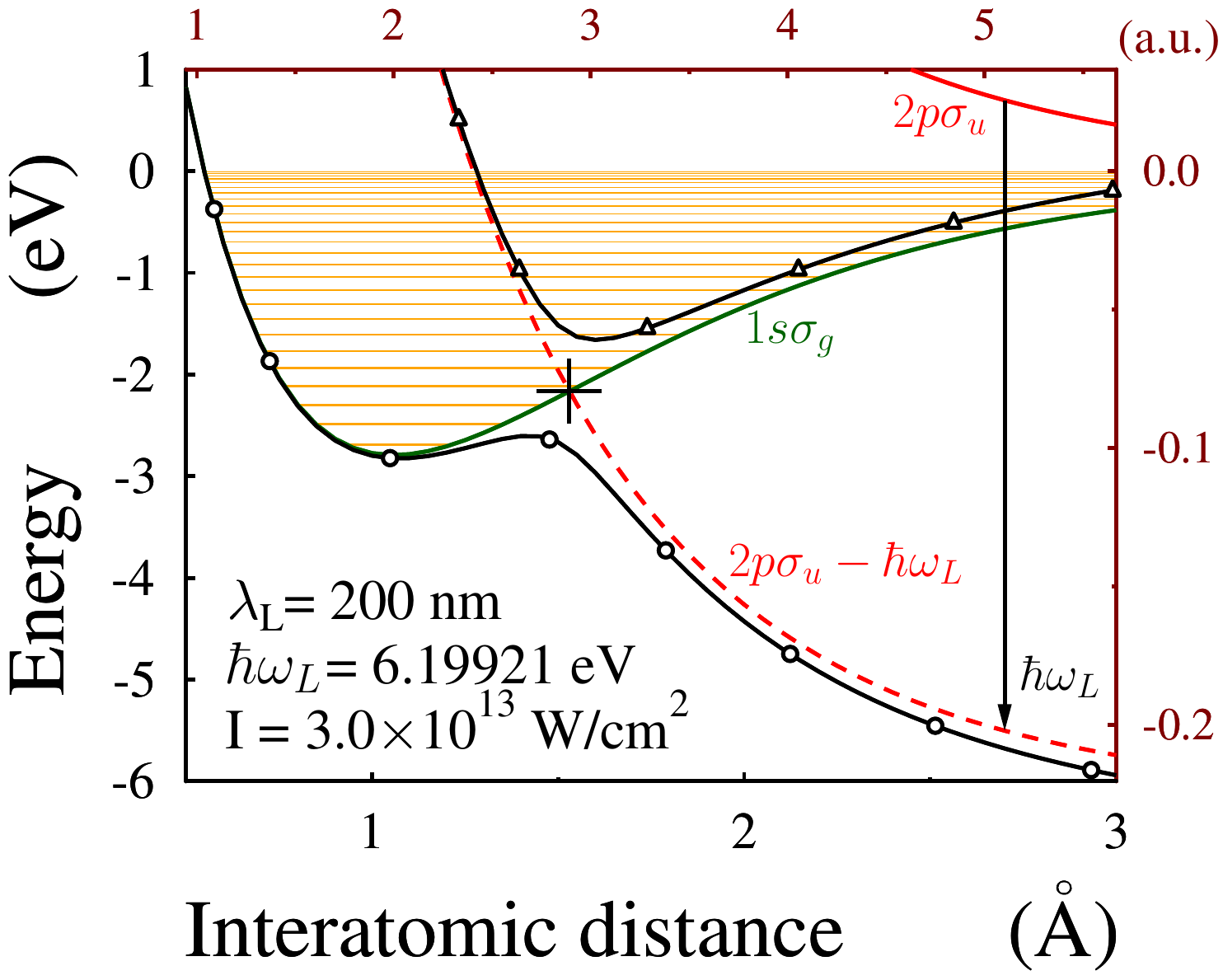}

\caption{\label{fig:1}Potential energies of the $\mathrm{D}_{2}^{+}$ molecule
and the light-induced conical intersection (LICI).\textbf{ Left panel:}
The dressed adiabatic surfaces as a function of the interatomic distance
$R$ and the angle $\theta$ between the molecular axis and the laser
polarization exhibiting the LICI for a field intensity of $3\times10^{13}\frac{W}{cm^{2}}$.
\textbf{Right panel:} The diabatic energies of the ground $\left(1s\sigma_{g}\right)$
and the first excited $\left(2p\sigma_{u}\right)$ states of the $\mathrm{D}_{2}^{+}$
molecule are displayed by solid green and red lines, respectively.
The field dressed excited state ($2p\sigma_{u}-\hbar\omega_{L}$;
dashed red line) forms a LICI with the ground state. A cut through
the adiabatic surfaces at $\theta=0$ (parallel to the field) is depicted
by solid black lines marked with circles (lower adiabatic state) and
triangles (upper adiabatic state). The position of the LICI is denoted
with a cross ($R_{LICI}=1.53\,\textrm{\AA}=2.891\, a.u.$ and $E_{LICI}=-2.166\, eV$).}
\end{figure*}

\textbf{Light-induced conical intersection (LICI).} In the dressed
representation the laser light shifts the energy of the $2p\sigma_{u}$
repulsive excited potential curve by $\hbar\omega_{L}$ and a crossing
between the ground $(V_{1})$ and the shifted excited $(V_{2}-\hbar\omega_{L})$
potential energy curves is created. By diagonalizing the potential
energy matrix \cite{Gabor1} one obtains the adiabatic potential surfaces
$V_{lower}$ and $V_{upper}$ (Fig. \ref{fig:1}). These two surfaces
cross each other at a single point $R$, $\theta$, giving rise to
a conical intersection whenever the conditions $\cos\theta=0$ $(\theta=\pi/2)$
and $V_{1}(R)=V_{2}(R)-\hbar\omega_{L}$ are simultaneously fulfilled
\cite{Milan1}. 

The characteristic properties of the LICI, i.e., the location of the
intersections and the strengths of the nonadiabatic couplings, can
be directly controlled by the laser frequency and intensity. This
opens the possibility to control the nonadiabatic effects emerging
from the LICIs. To demonstrate the impact of the LICI on the photodissociation
dynamics of D$_{2}^{+}$ we have to solve the time-dependent nuclear
Schrödinger equation (TDSE) using the Hamiltonian $\hat{H}$ given
by Eq. (\ref{eq:Hamilton}). The angular distribution of the photofragments
$P(\theta)$ were calculated with the solution of the TDSE equation.

\textbf{Methods.} The MCTDH (multi configuration time-dependent Hartree)
method was used to solve the TDSE \cite{Dieter1,Dieter2,Dieter3,Dieter4,Dieter5}.
It is one of the most efficient approaches for solving the TDSE. For
describing the vibrational degree of freedom we have applied FFT-DVR
(Fast Fourier Transformation-Discrete Variable Representation) with
$N_{R}$ basis elements distributed within the range from 0.1 a.u.
to 80 a.u. for the internuclear separation. The rotational degree
of freedom was represented by Legendre polynomials $\left\{ P_{J}(\cos\theta)\right\} _{j=0,1,2,\cdots,N_{\theta}}$.
These so called primitive basis sets ($\chi$) were used to build
up the single particle functions ($\phi$), which in turn were applied
to represent the wave function:
\begin{eqnarray}
\phi_{j_{q}}^{(q)}(q,t) & = & \sum_{l=1}^{N_{q}}c_{j_{q}l}^{(q)}(t)\;\chi_{l}^{(q)}(q)\qquad q=R,\,\theta\label{eq:MCTDH-wf-1}\\
\psi(R,\theta,t) & = & \sum_{j_{R}=1}^{n_{R}}\sum_{j_{\theta}=1}^{n_{\theta}}A_{j_{R},j_{\theta}}(t)\phi_{j_{R}}^{(R)}(R,t)\phi_{j_{\theta}}^{(\theta)}(\theta,t).\nonumber 
\end{eqnarray}
In the actual simulations $N_{R}=2048$ and $N_{\theta}=199$ were
used. On both diabatic surfaces and for both degrees of freedom a
set of $n_{R}=n_{\theta}=20$ single particle functions were applied
to construct the nuclear wave packet of the system. The calculations
converged properly by using these chosen parameters. Using the solution
of equation (\ref{eq:MCTDH-wf-1}) one can calculate the angular distribution
of the photofragments \cite{Dieter3}:
\begin{equation}
P(\theta_{j})=\frac{1}{w_{j}}\intop_{0}^{\infty}dt<\psi(t)|W_{\theta_{j}}|\psi(t)>\label{eq:angdist-1}
\end{equation}
where $-iW_{\theta_{j}}$ is the projection of the complex absorbing
potential (CAP) on a specific point of the angular grid $\left(j=0,..N_{\theta}\right)$,
and $w_{j}$ is the weight related to this grid point according to
the applied DVR.

\textbf{The simulations}. Linearly polarized Gaussian laser pulses
centered around $t=0\, fs$ were applied in the numerical calculations
with a carrier wavelength of $\lambda_{L}=200\, nm$. The pulse duration
at full width of half maximum (FWHM) is $t_{pulse}=30\, fs$. The
initial nuclear wave packet (at $t=-100\, fs$) was chosen to be in
its rotational ground state ($J=0$) and in one of its vibrational
eigenstates $\left(\nu=4,5,6,7\right)$. Under such conditions, the
angular distribution of the photofragments can provide accurate details
of the photodissociation of single vibrational levels \cite{Sanding}. 

Full two dimensional calculations (2d) in $R-\theta$ space have been
performed. In order to have a better understanding of the obtained
results, we have also performed calculations with a restricted one
dimensional (1d) approach where we have eliminated the rotational
motion from the system by putting $L_{\theta\varphi}$ to zero in
the Hamiltonian Eq. (\ref{eq:Hamilton}). This way the molecule's
initial orientation can not change during the dissociation process,
and the TDSE can be solved with independent 1d calculations for each
value of $\theta$ using the ``effective field strength'' $\epsilon_{0}^{eff}=\epsilon_{0}\cdot\cos\theta$
$($intensity $I_{0}^{eff}=I_{0}\cdot\cos^{2}\theta)$ at that value
of $\theta$. In these 1d calculations the Hamiltonian depends only
parametrically on the rotational degree of freedom $\left(\theta\right)$
and none of the individual calculations are able to take into account
the effects of the LICI. The respective potential energy curves exhibit
avoided crossings as can be seen for $\theta=0$ in Fig. \ref{fig:1}.
\begin{figure*}[p]
\includegraphics[width=0.4\paperwidth]{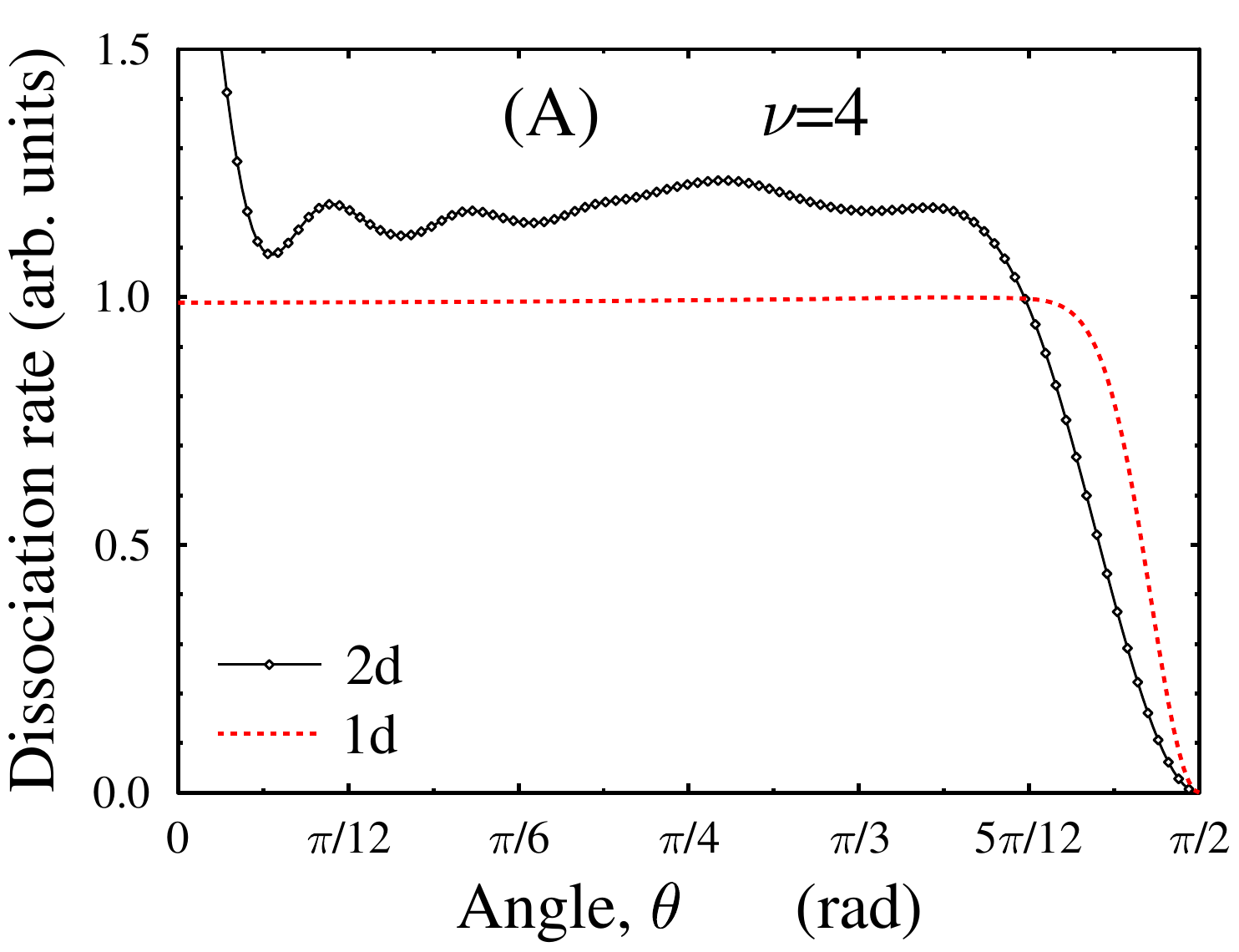}\includegraphics[width=0.4\paperwidth]{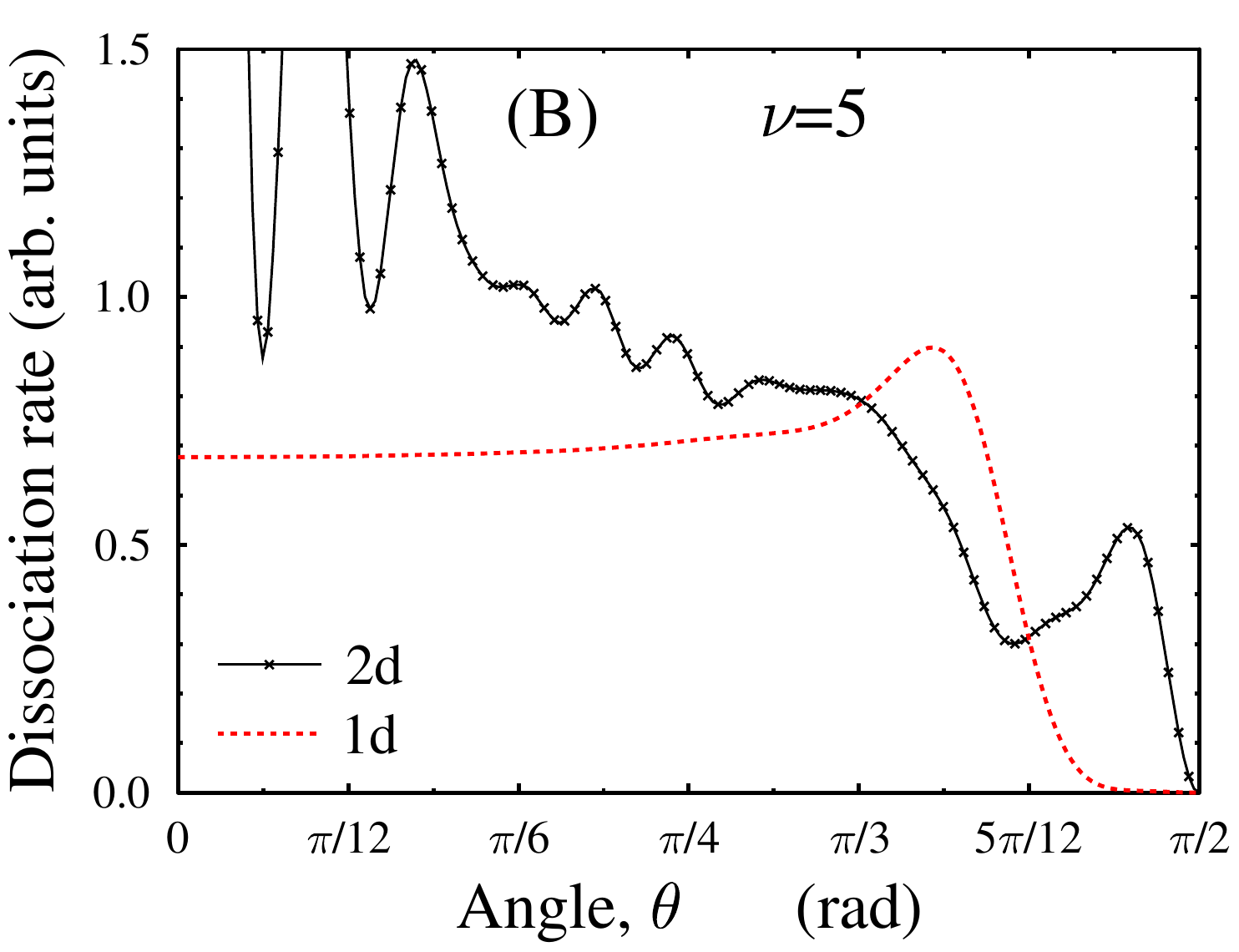}

\includegraphics[width=0.4\paperwidth]{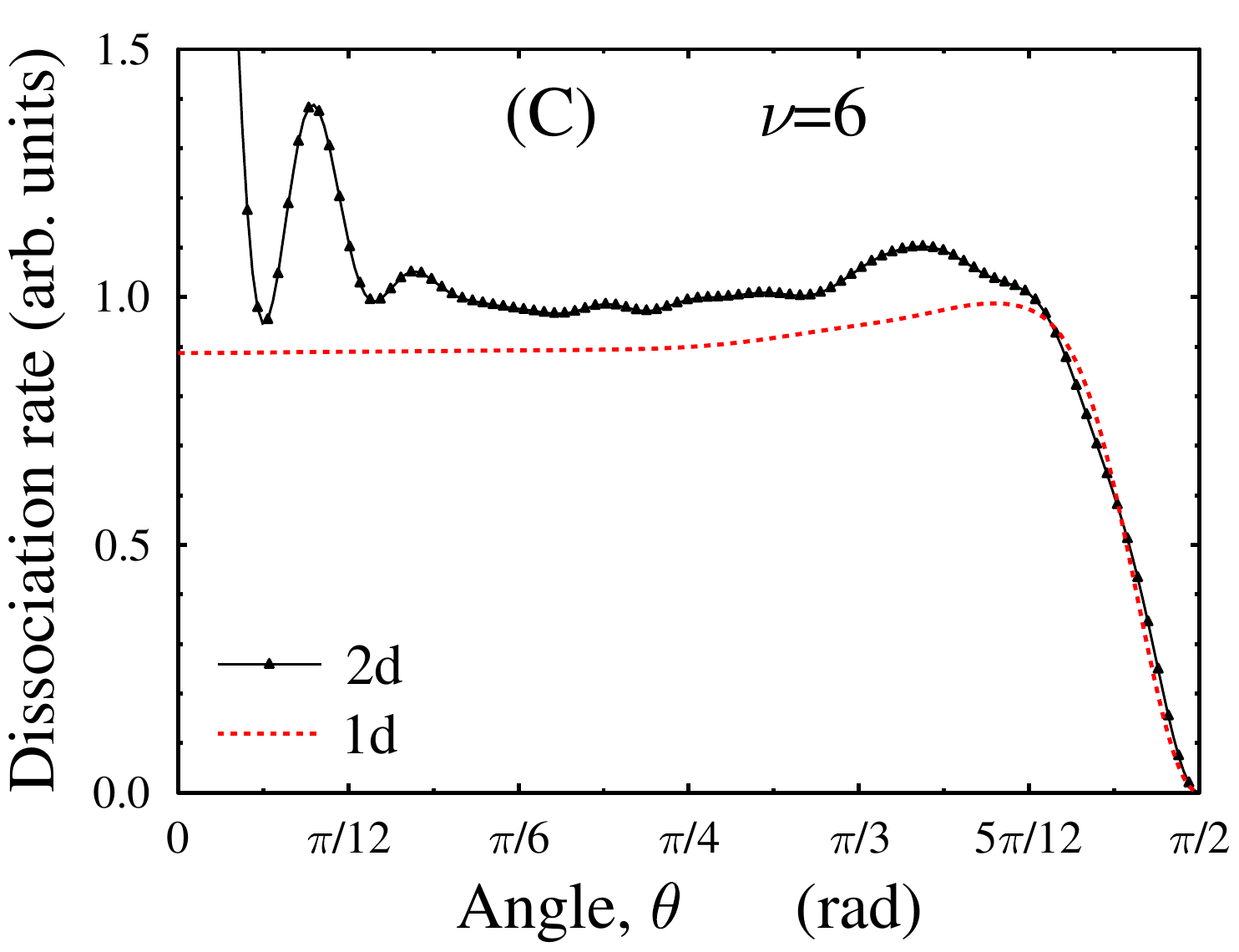}\includegraphics[width=0.4\paperwidth]{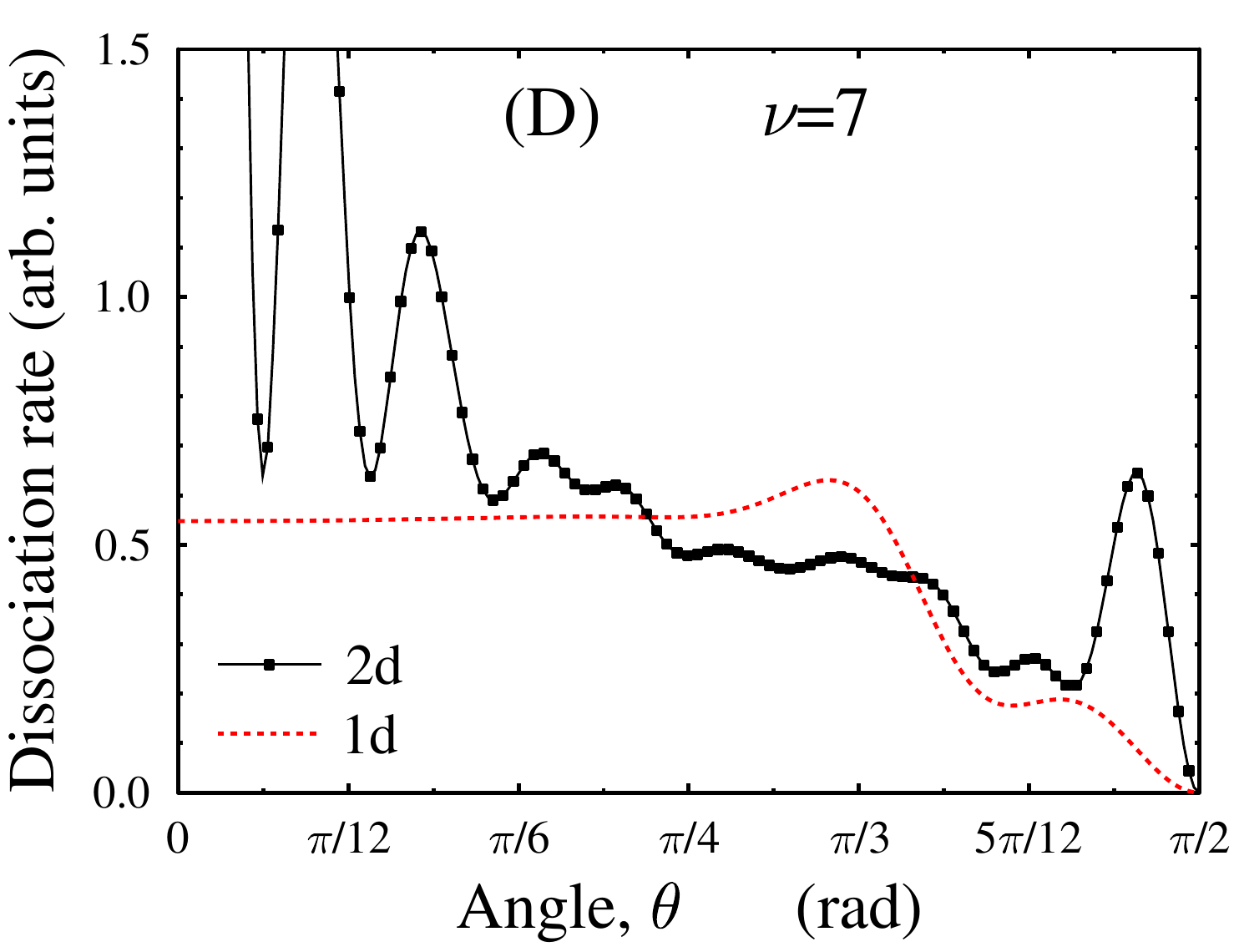}

\caption{\label{fig:2}Fragment angular distributions of the dissociating $\mathrm{D}_{2}^{+}$
molecule for four different initial vibrational states ($\nu=4,5,6,7$).
Curves are presented both for one dimensional (1d) and for two dimensional
(2d) cases. The applied field intensity is $1\times10^{14}\, W/cm^{2}$.}
\end{figure*}
\begin{figure*}[p]
\begin{centering}
\begin{minipage}[c]{0.9\textwidth}%
\begin{center}
\includegraphics[width=0.32\textwidth]{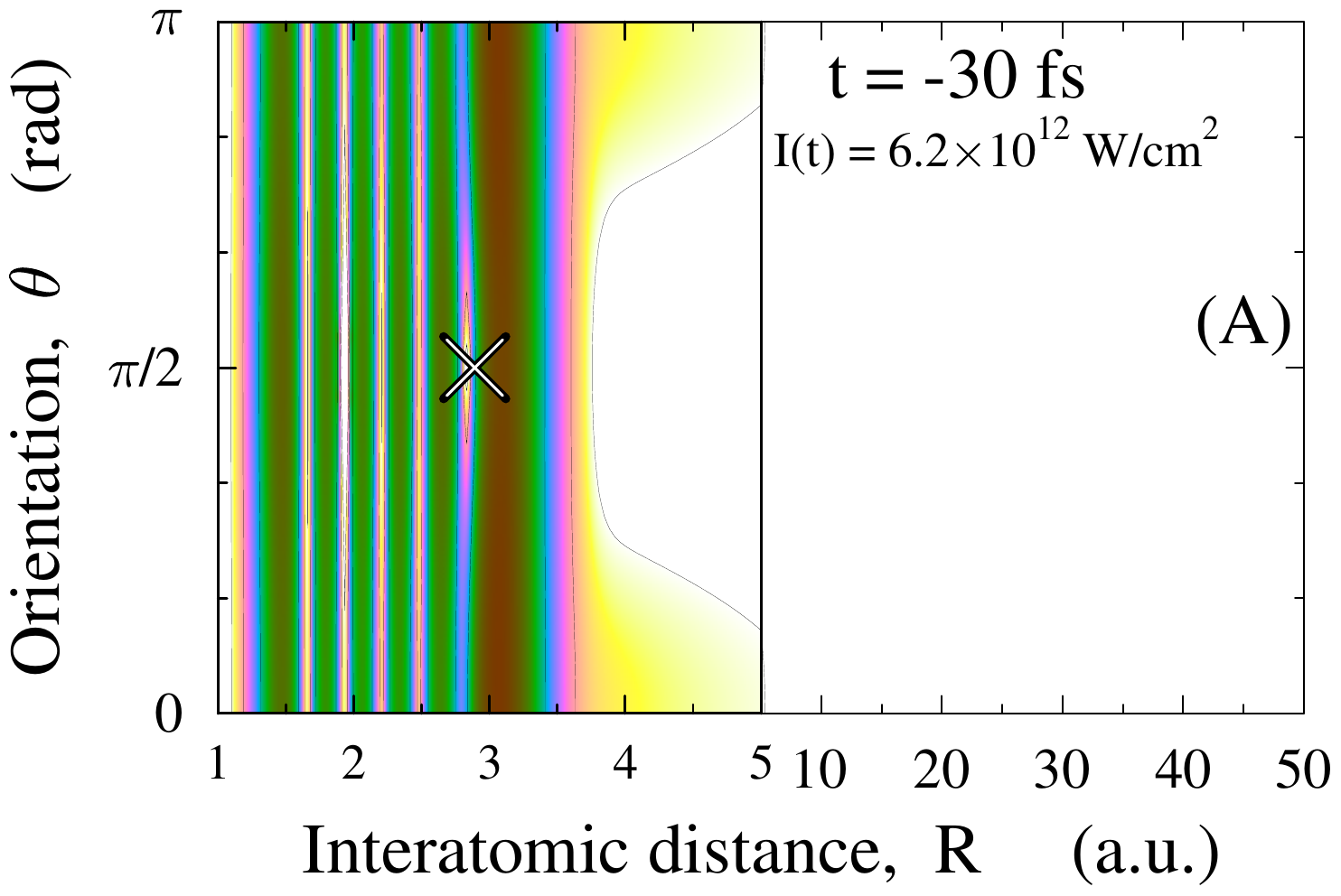}\hfill{}\includegraphics[width=0.32\textwidth]{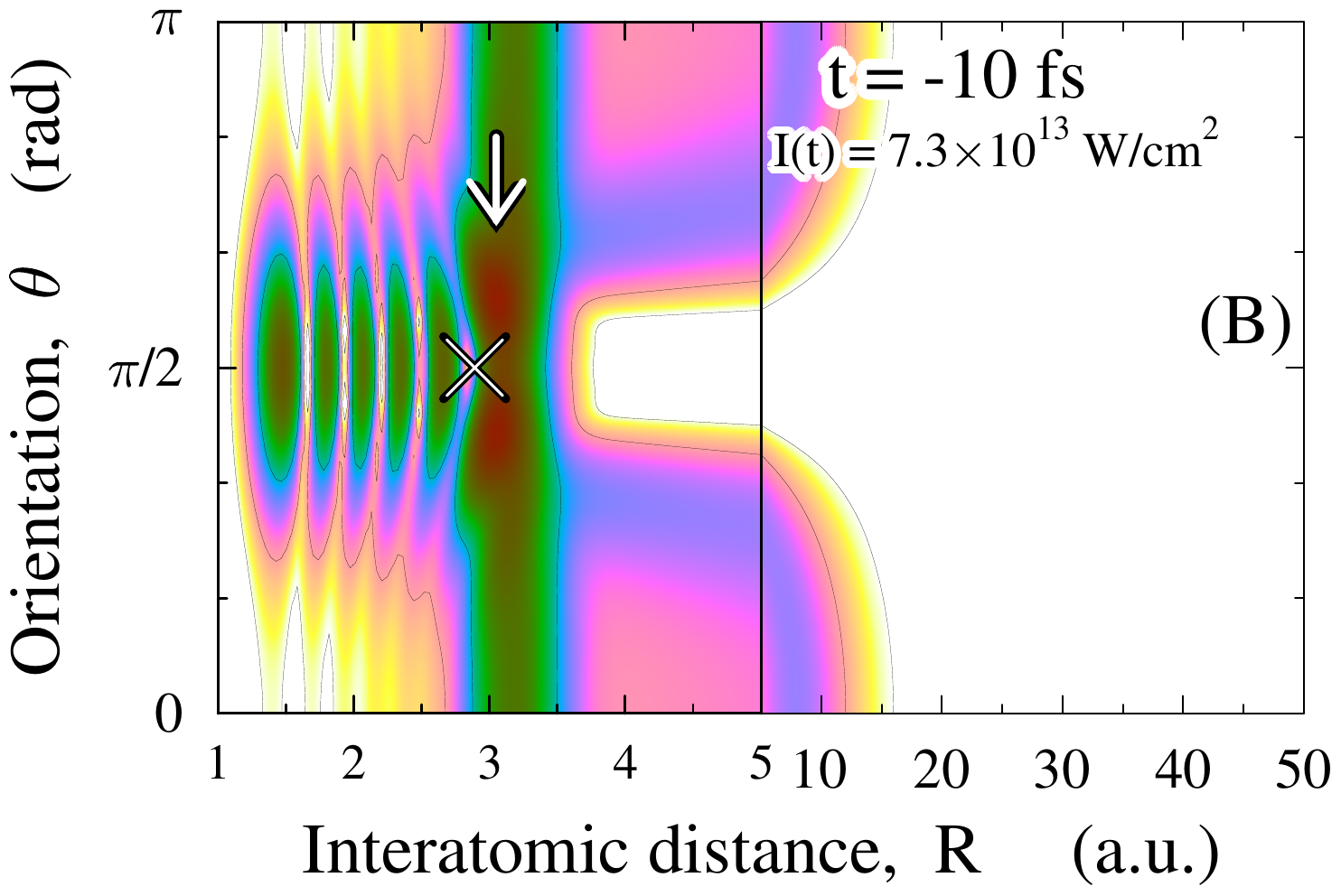}\hfill{}\includegraphics[width=0.32\textwidth]{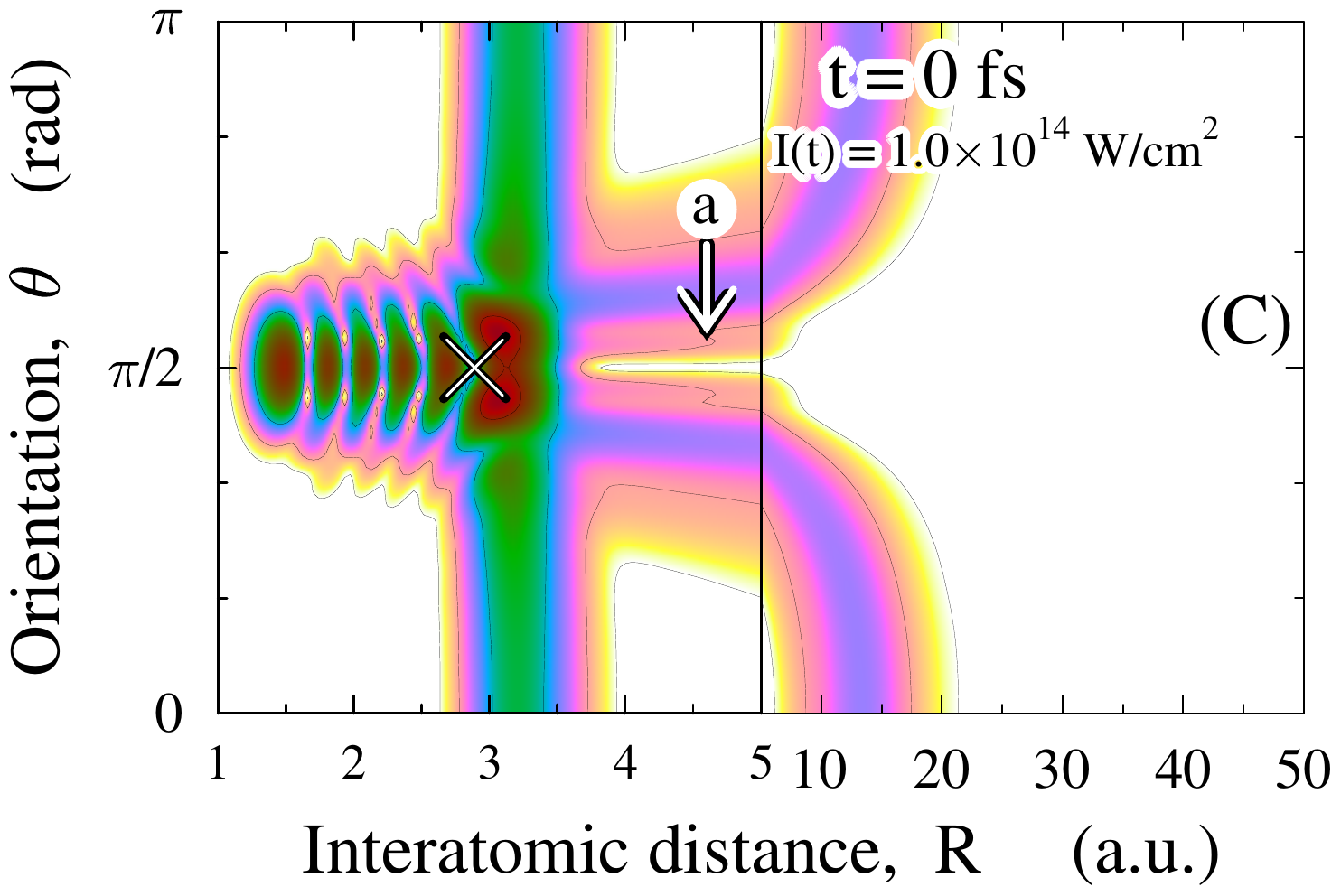}
\par\end{center}

\begin{center}
\includegraphics[width=0.32\textwidth]{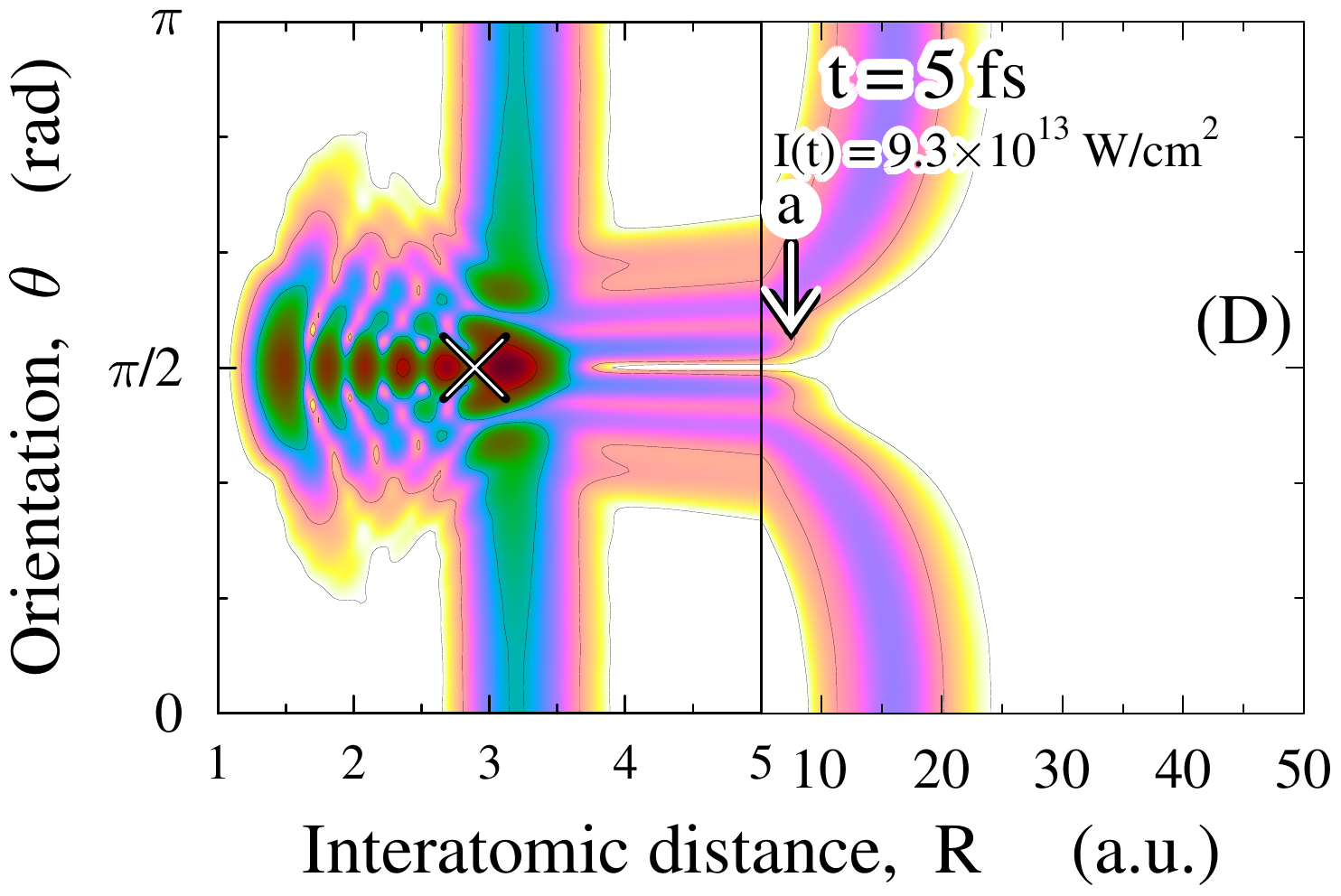}\hfill{}\includegraphics[width=0.32\textwidth]{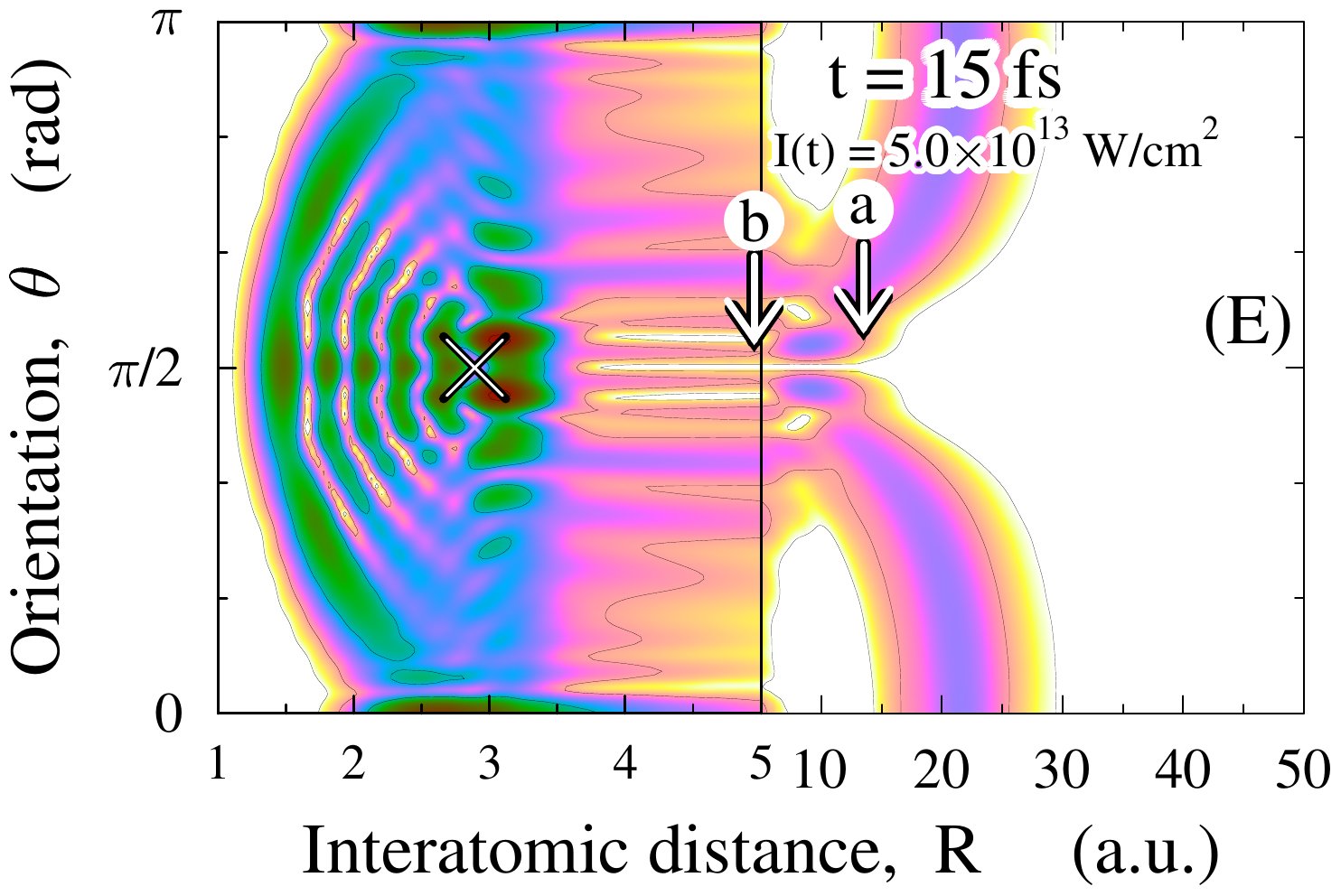}\hfill{}\includegraphics[width=0.32\textwidth]{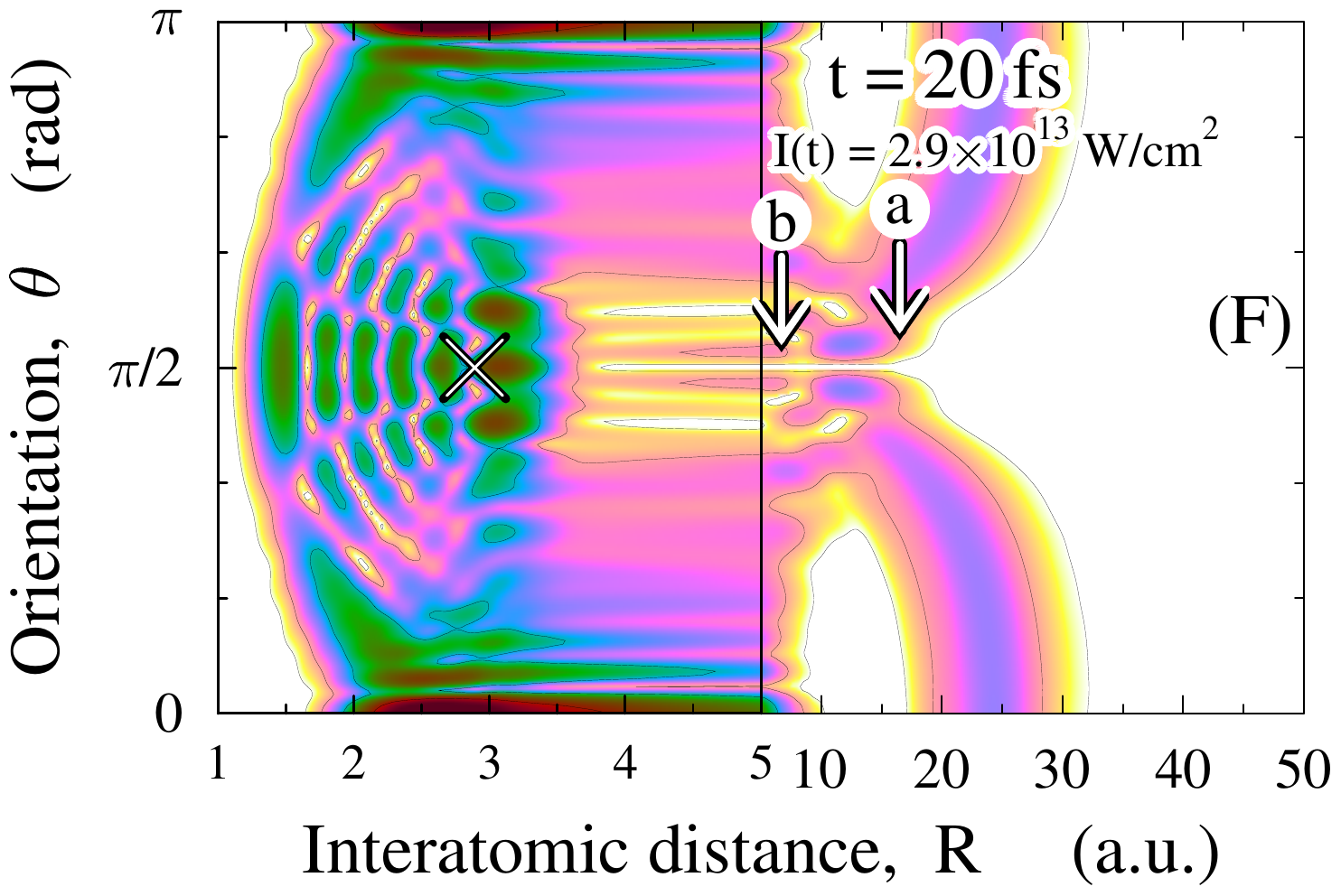}
\par\end{center}

\begin{center}
\includegraphics[width=0.32\textwidth]{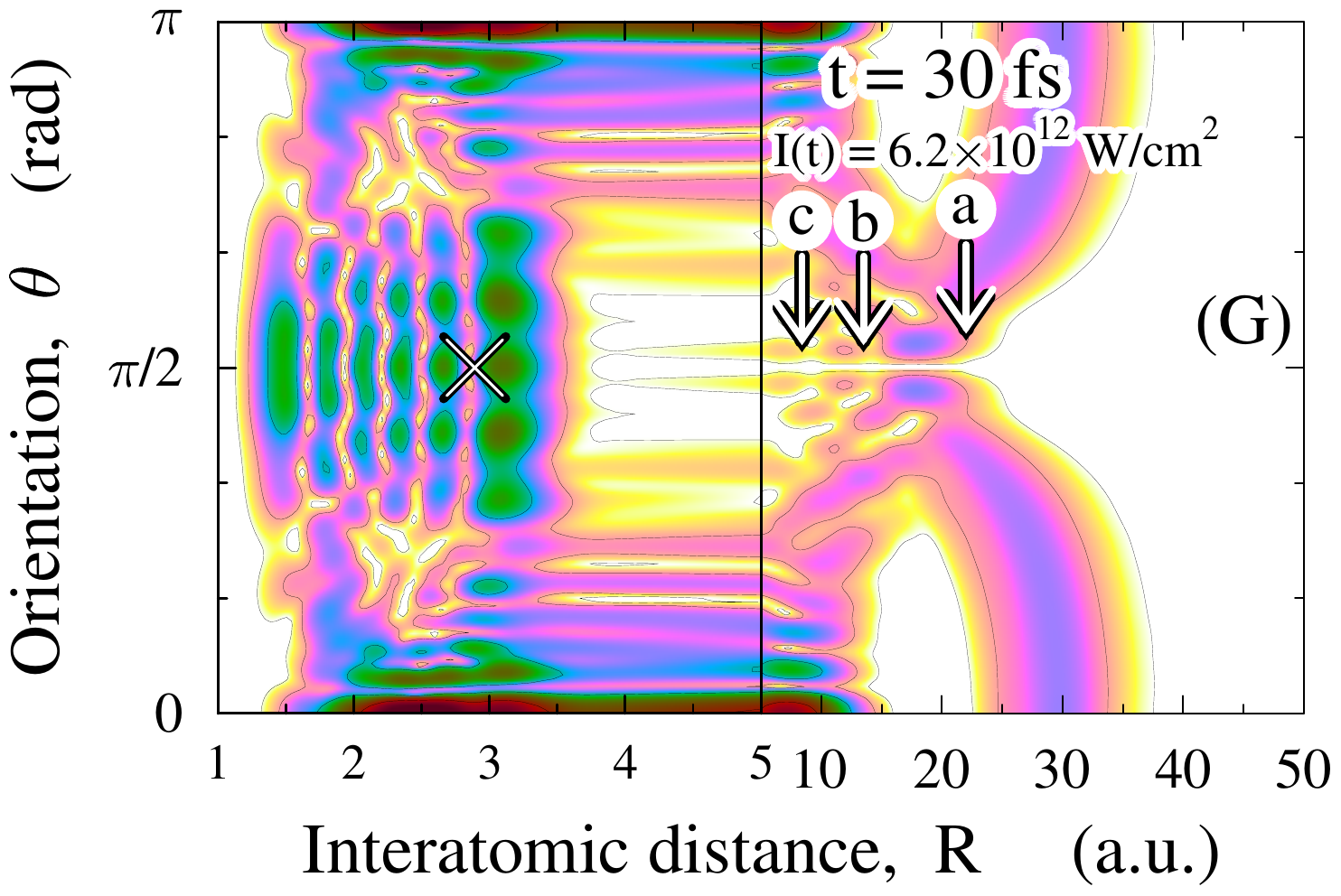}\hfill{}\includegraphics[width=0.32\textwidth]{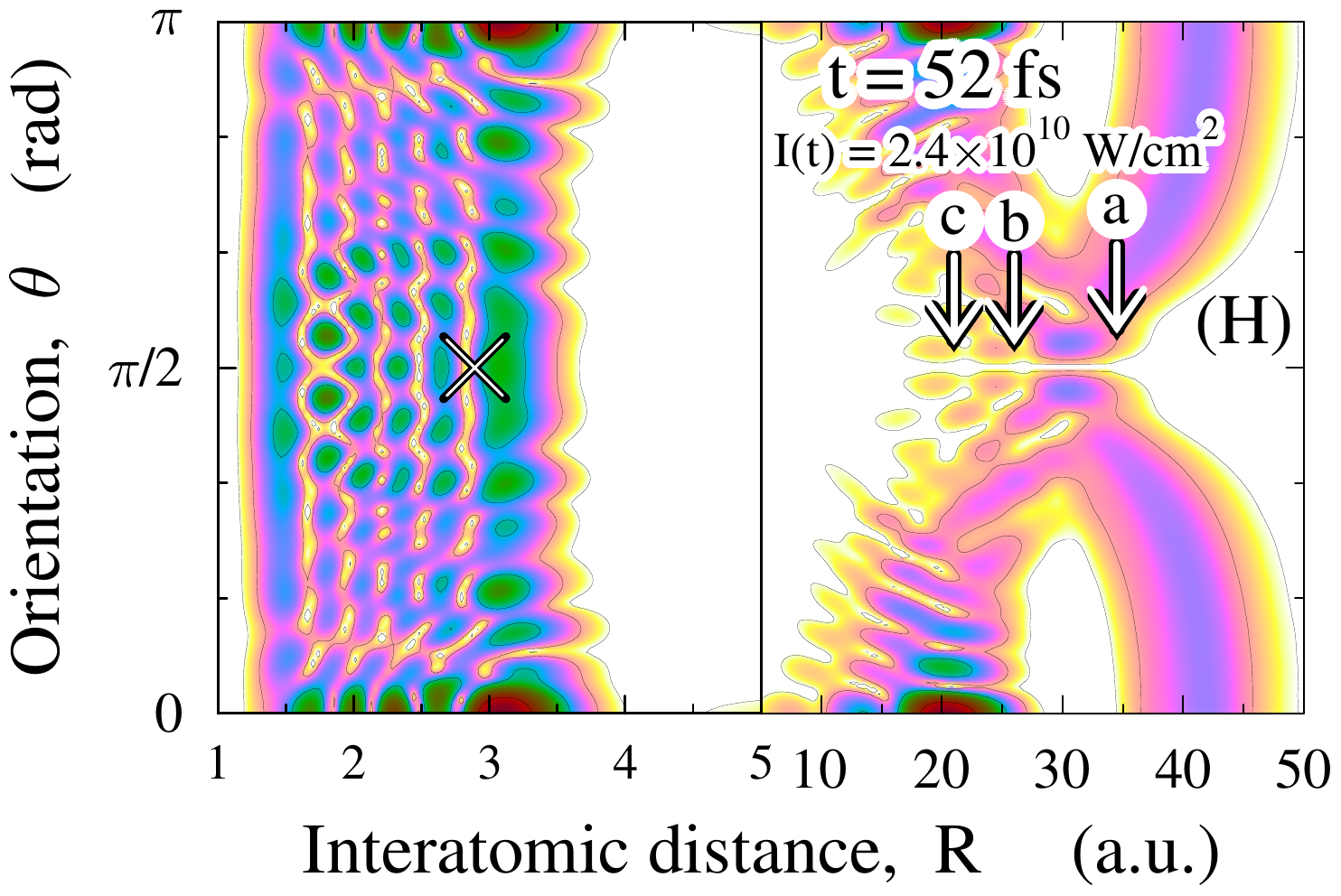}\hfill{}\includegraphics[width=0.32\textwidth]{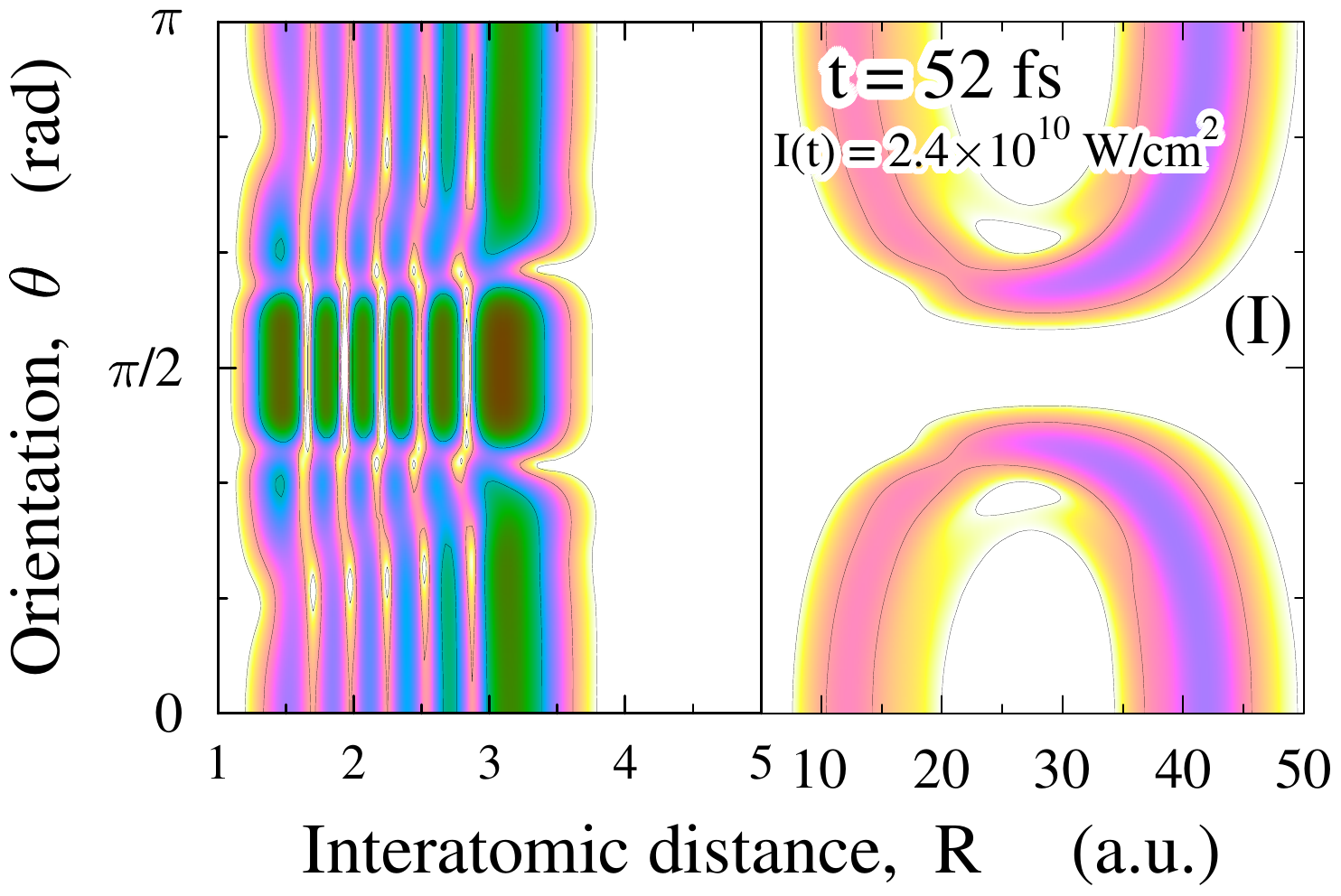}
\par\end{center}%
\end{minipage}\hfill{}%
\begin{minipage}[c]{0.09\textwidth}%
\begin{center}
\includegraphics[width=1\textwidth]{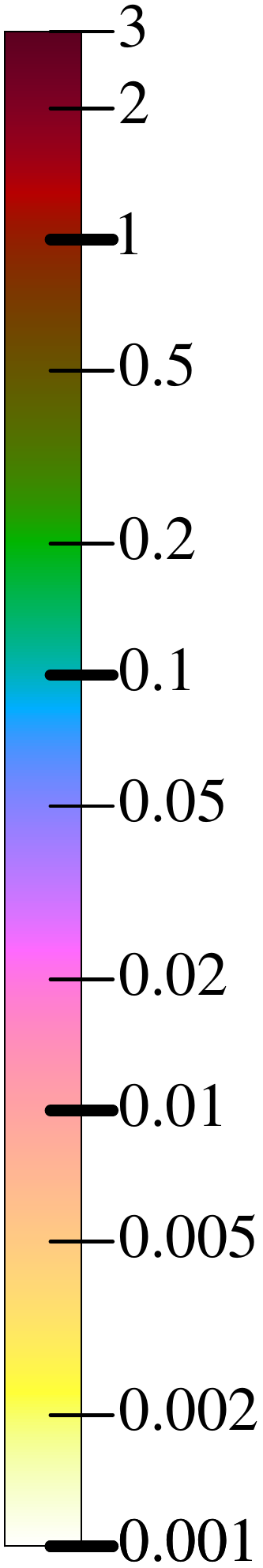}
\par\end{center}%
\end{minipage}
\par\end{centering}

\caption{\label{fig:3}Snapshots from the real-time evolution of the nuclear
density of the $\nu=5$ vibrational state of $\mathrm{D}_{2}^{+}$
due to a Gaussian laser pulse of peak intensity $1\times10^{14}W/cm^{2}$
and $30\, fs$ duration. The nuclear density exhibits severe interference
effects and splits at larger distances around $\theta=\pi/2$. The
instantaneous intensity is shown in the individual snapshots. The
yellow cross denotes the position of the LICI. The points a, b and
c are explained in the text. All panels show the results of the full
2d calculation expect of the last panel (I) which is computed in 1d
for comparison. Note the jump in the interatomic scale. }
\end{figure*}

The results obtained for the angular distribution of the dissociation
rate for the above described pulse with a peak field intensity of
$1\times10{}^{14}\frac{W}{cm^{2}}$ are displayed in Figs. \ref{fig:2}a-\ref{fig:2}d
for the four different initial vibrational levels $\nu=4-7$. The
scale on Fig. \ref{fig:2} was chosen such that the dissociation rate
of 1 implies that the dissociation in a given direction is complete.
Consequently, in 1d the dissociation rate can not be larger than 1.
Larger values in the full 2d calculation mean that some parts of the
dissociating particles were rotated by the field to this direction
from some different initial directions.

Let us study first the 1d curves in Fig. \ref{fig:2}. For the case
of $\nu=4$ vibrational level the dissociation rate decreases monotonically
as a function of $\theta$ and is very close to the value of 1 up
to $\theta=5\pi/12$. This is related to the fact that the energy
of this vibrational level is just above the energy of the LICI and,
therefore, most of the initial wavepacket belongs to the lower dissociative
adiabatic surface and can easily dissociate. For larger $\nu$ the
1d dissociation rate curves display one or two ($\nu=7$) local maxima
as a function of the angle $\theta$. This behavior is related to
the accidental increase of the so - called bond hardening effect \cite{Bandrauk5}
at particular wavelengths and intensities of the laser field \cite{Bandrauk3}.
The 1d dissociation rate predicts a significant bond hardening effect
for $\nu=5$ and $\nu=7$, even at the parallel orientation of the
molecules. In a Gaussian laser pulse the field rises smoothly and
independently of the peak intensity starts to interact with the molecular
system at weak intensities, and therefore the initial populations
on the lower and on the upper adiabatic surfaces are determined by
the behavior of the system at low intensities. In our earlier work
\cite{Gabor7}, it was shown for the $\nu=5$ and $\nu=7$ initial
vibrational levels that the $\lambda_{L}=200\, nm$ wavelength leads
to a close to an ``ideal'' bond hardening case at weak intensities.
This is the reason why we find here an significant bond hardening
effect up to $1\times10^{14}\frac{W}{cm^{2}}$ field intensities at
this wavelength.

Let us now turn to the 2d results. For the initial $\nu=4$ vibrational
eigenstate, the 2d curve (see Fig. \ref{fig:2}a) is related to the
1d one as one can expect it from the well known light induced potential
picture \cite{Charron1}. The differences arise due to our finding
that the laser light not just drives the dissociation process, but
also starts to rotate the molecules. As a result the dissociation
rate decreases at large $\theta$ angles and at the same time increases
at the small values of $\theta$. As the induced rotation towards
the polarization direction of the electric field creates rotational
nodes, the angular distribution of the dissociation rate is further
affected by the appearance of additional structures \cite{Gabor6}.

For the $\nu=6$ vibrational level (see Fig. \ref{fig:2}c) the increase
of the dissociation rate at small angles is not accompanied by its
decrease at orientations close to the direction perpendicular to the
polarization axis. Actually, very close to $\theta=\pi/2$ the dissociation
rate in 2d is slightly larger than in 1d. This may be interpreted
as follows. The population of the upper adiabatic surface is rotated
by the electric field into this direction and later on this extra
population can dissociate by population transfer to the lower adiabatic
surface in the vicinity of the LICI. 

From this explanation we may expect more pronounced effects for the
$\nu=5$ and $\nu=7$ cases, where the bond hardening is more effective
as concluded from the study in 1d. Indeed, in these cases the 2d dissociation
rates in Fig. \ref{fig:2} exhibit large peaks around $\theta\gtrsim11\pi/24$,
whose heights are one ($\nu=7$) or even two ($\nu=5$) orders of
magnitude larger than the corresponding dissociation rates in the
1d calculations. This characteristic structure of the 2d curves in
this angle region can not be explained without the strong nonadiabatic
effect due to the existence of the LICI. The peaks displayed on the
2d curves are direct fingerprints of the existence of the laser-induced
conical intersection in the studied system. We note here that such
peaks can also be seen at smaller intensities (1$\times$10$^{13}\frac{W}{cm^{2}}$,
3$\times$10$^{13}\frac{W}{cm^{2}}$), but in less pronounced form.

To understand the underlying dynamical process more deeply, we analyzed
the nuclear density function $\left|\psi(R,\theta,t)\right|^{2}\left(=\left|\psi^{1s\sigma_{g}}(R,\theta,t)\right|^{2}+\left|\psi^{2p\sigma_{u}}(R,\theta,t)\right|^{2}\right)$.
Snapshots of the nuclear wave packet density functions are shown in
Fig. \ref{fig:3}. After the pulse is over ($t=52\, fs$, Fig. \ref{fig:3}h)
we can identify three peaks -- on both sides of the symmetry axis
$\theta=\pi/2$ -- in the region of interest $11\pi/24\lesssim\theta\lesssim13\pi/24$
of the outgoing wavepacket. These peaks -- especially the most right
one -- are clearly responsible for the characteristic structure of
the huge excess in the dissociation probability close to $\theta=\pi/2$.
Their ``leading edge'' are labeled by ``a'', ``b'' and ``c''
on the subfigures whenever they are recognizable. 

In what follows we study the time evaluation of the nuclear density
in order to see the origin of these peaks. It can be seen that at
$t=-30\, fs$ (Fig. \ref{fig:3}a) the molecules start to dissociate
on the lower adiabatic surface. This process evolves in time, but
at $t=-10\, fs$ one can clearly recognize also the fingerprint of
the rotational motion induced on the upper surface (see the arrow
in Fig. \ref{fig:3}b). At $t=0\, fs$ (Fig. \ref{fig:3}c, labeled
by ``a'') tracks show up that an additional dissociation process
starts close to $\theta=\pi/2$, presumably from the upper adiabatic
surface. From $t=15\, fs$ (Fig. \ref{fig:3}e) on it is clearly seen
that additional dissociation takes place continuously, and from $t=20\, fs$
(Fig. \ref{fig:3}f) on the dissociated fragments explicitly appear
in the dissociation region ($R>5a.u.$). The much less populated ``b''
and ``c'' peaks appear for the first time on subfigures Fig. \ref{fig:3}e
and Fig. \ref{fig:3}g, respectively.

The last snapshot displays the results of the 1d simulation at $t=52\, fs$,
the same time as in Fig. \ref{fig:3}h for the full 2d calculation.
The dynamics during the initial time period, when the instantaneous
intensity is low, are very similar in both the 1d and 2d models. After
reaching the maximum intensity, however, two significant differences
can be observed. The first one is, as expected, the unequivocal fingerprint
of the presently discussed nonadiabatic effect leading to a strong
enhancement of fragment density near $\theta=\pi/2$ in the 2d model,
while the second one is the appearance of interference patterns in
the 2d model. As both effects rely on the dissociation of those molecules
which have been previously rotated by the field, the 1d model is unable
to account for them. On the other hand, it is not surprising that
at the beginning of the dissociation process, when the rotation develops
slowly, the 1d and 2d approaches lead to the similar smooth behavior
in the dissociation region.

In summary, the results obtained clearly show the direct impact of
the laser-induced conical intersection on the dissociation dynamics
of the D$_{2}^{+}$ molecule. The structure and magnitude of the 2d
($\nu=5,7$) dissociation rates close to $\theta=\pi/2$ undoubtedly
demonstrate the strong nonadiabatic effects due to the presence of
the LICI. The strong nonadiabaticity turns the molecules perpendicular
to the polarization direction and then these molecules can travel
through the LICI to the lower adiabatic surface on which they dissociate.
This process can only happen if population transfer takes place via
the laser-induced conical intersection.

\section*{Acknowledgements}

The authors acknowledge financial support by the Deutsche Forschungsgemeinschaft
(Project ID CE10/50-2). Á. V. also acknowledges the TÁMOP-4.2.4.A/
2-11/1-2012-0001 `National Excellence Program' and the OTKA (NN103251)
project. This research was supported in part by the National Science
Foundation under Grant No. NSF PHY11-25915.

\end{document}